\documentstyle[epsfig,aps]{revtex}
  
\begin{document}    
\title{Study of scalar gradient fields by geometric measure theory}
\author{J\"org Schumacher}  
\address{Fachbereich Physik, Philipps-Universit\"at, D-35032 Marburg, Germany} 
\date{\today}
\maketitle  

\begin{abstract}
Upper bounds of the Hausdorff volume of scalar gradient field graphs
are derived by means of geometric measure theory.  The approach reproduces that
scalar gradient fields along a mean imposed scalar gradient become space filling
for sufficiently high values of Schmidt numbers $Sc$.  The bounds are consistent
with findings from recent high-resolution numerical experiments for $1\le Sc\le 64$,
but too rough when compared with numerical simulations.  
A Reynolds number dependence of the bounds
is found due to the additional scalar gradient
stretching term in the equation of motion.

\end{abstract}

\vspace{1cm}
\noindent
Recent direct numerical simulations (DNS) suggested that the passive scalar
mixing in a turbulent flow becomes more isotropic when the Schmidt number,
$Sc=\nu/\kappa$, is increased to values larger than unity, but the Taylor
Reynolds number, $R_{\lambda}$, of the advecting turbulent flow is kept constant
\cite{PK02,Brethouwer}.  Here $\nu$ is the kinematic viscosity of the fluid and
$\kappa$ the diffusivity of the passive scalar field $\theta({\bf x},t)$.  The
scalar was driven by a mean scalar gradient, ${\bf G}=G{\bf e}_{\alpha}$, in
both simulations which causes deviations from isotropy of the small-scale
statistics.  For increasing $Sc$, scalar filaments can be advected to ever finer
scales which steepens up the local gradients, $g_i=\partial_i \theta$.  Large
gradients (or fronts) that are aligned with the mean, i.e.  $g_{\alpha}$, occur
already for $Sc\sim 1$ or even below and are associated with characteristic
scalar structures, so-called ramps and cliffs (for further references, see
\cite{Warhaft}).  A return to isotropic mixing is then thought as a growing
compensation of those pronounced positive fronts by an increasing number of
steep negative gradients with increasing $Sc$.  Consequently, the probability
density function of $g_{\alpha}$ gets more symmetric tails and odd order
derivative moments decay.

Our understanding of this process for advection in Navier-Stokes flows is still
incomplete and an investigation of geometric properties of scalar gradient
fields was started therefore recently \cite{Schu03}.  Fractal and multifractal
properties of scalar gradient level sets and related quantities were studied
there in a series of high-resolution DNS where, e.g., a higher degree of local
isotropy was found to be related to a flater spectrum of generalized
(multifractal) dimensions.

On the theoretical side, concepts of geometric measure theory \cite{Mor88} were
used sucessfully for scalars in turbulence \cite{Con91,Con93} and relations to
the scaling behavior of low-order structure functions were established.  An
extension of the framework discussed the dependence of the geometric properties
of scalar level sets on Schmidt (or Prandtl) number and on spatial separation
scales, respectively \cite{GroLoh94}.  The purpose of the present brief report
is to step in at this point and to extend the approach to scalar gradient
fields.  We will focus on the physics in the Batchelor regime of scalar
turbulence where the advecting flow is in its viscous subrange \cite{Batchelor}.
Scales below the Kolmogorov dissipation scale of fluid turbulence,
$\eta=(\nu^3/\epsilon)^{1/4}$, but larger than the Batchelor scale
$\eta_B=\eta/Sc^{1/2}$ are considered.  $\epsilon$ is the energy dissipation
rate of the flow.  We will derive an upper bound for the scaling dimension of
the Hausdorff volume of the scalar gradient field graph.  This can give us an
idea of how the gradients are distributed spatially and how this distribution
depends on $Sc$.  Our findings will be compared finally with high resolution
numerical data for Schmidt numbers between 1 and 64 to test the sensitivity of
the derived scaling dimension bounds.

The geometric measure theory generalizes concepts of differential geometry to
non-smooth hypersurfaces embedded in an Euclidian space.
The central object is the graph of the field under consideration, which 
is defined as
%--------------------------------------------------------------- 
\begin{eqnarray}
\Gamma_r=\{({\bf x},g_{\alpha})|{\bf x}\in
B_r\subset \mbox{R}^3,\, g_{\alpha}=g_{\alpha}({\bf x})\}\,.
\end{eqnarray}
%---------------------------------------------------------------
In case of a smooth field this is a three-dimensional hypersurface embedded
in four-dimensional space.
In Fig.~1 such graph is shown over a two-dimensional plane that is
vertical to the direction of ${\bf G}$. 
We observe that the hypersurface is strongly folded and rough for larger scales
thus suggesting fractal properties.

The Hausdorff dimension of such a graph, $D_H$, 
(or more precisely the box counting dimension) 
is obtained from the scaling behavior of its Hausdorff volume 
$H(\Gamma_r)$ with respect to scale $r$, the radius of balls $B_r$ of the
covering \cite{Mor88,Fal85},
%--------------------------------------------------------------  
\begin{eqnarray}  
\label{scal}  
H(\Gamma_r)=\int_{B_r}\,
\sqrt{1+r^2|{\bf\nabla}\tilde{g}_{\alpha}|^2}\,\mbox{d}^3{\bf x}
\sim r^{D_H}\;.
\end{eqnarray}  
%--------------------------------------------------------------  
The relative Hausdorff volume $H(\Gamma_r)/V_r$ can be estimated as
%--------------------------------------------------------------  
\begin{eqnarray}  
\label{vol}  
\frac{H(\Gamma_r)}{V_r}
\le\sqrt{1+\frac{3}{4\pi r} \int_{B_r}\,|{\bf\nabla}\tilde{g}_{\alpha}|^2
\,\mbox{d}^3{\bf x}}\;,  
\end{eqnarray}  
%--------------------------------------------------------------  
where the Cauchy-Schwartz inequality was used 
and $V_r=4\pi\,r^3/3$.
The scalar gradient field $g_{\alpha}({\bf x},t)$ is measured in units of 
its root mean square (rms) value, i.e. 
$\tilde{g}_{\alpha}=g_{\alpha}/\sqrt{\langle g^2_{\alpha}\rangle_V}$.
Here $V=L^3$ where $L$ is the outer scale of turbulence. Similarily, 
$\tilde{G}=G/\sqrt{\langle g^2_{\alpha}\rangle_V}$. 

Progress is made now with a substitution of the gradient term in (\ref{vol}) 
by means of the advection-diffusion equation for the scalar gradient component,
%--------------------------------------------------------------
\begin{eqnarray}
\left[\partial_t+u_i \partial_i -\kappa \partial^2_i\right]
\tilde{g}_{\alpha} +
(\partial_{\alpha}u_i) \tilde{g}_i
=-\partial_{\alpha}u_{\alpha} \tilde{G}\,.
\end{eqnarray}
%--------------------------------------------------------------
From the resulting second order balance one gets
%--------------------------------------------------------------
\begin{eqnarray}
|{\bf\nabla}\tilde{g}_{\alpha}|^2&=&
\frac{1}{2\kappa} [(-u_i \partial_i+\kappa \partial_i^2)
\tilde{g}_{\alpha}^2\nonumber\\
&-&2\tilde{g}_{\alpha}(\partial_{\alpha}u_i \tilde{g}_i+
\partial_{\alpha}u_{\alpha}\tilde{G})]\,,
\label{substitution}
\end{eqnarray}
%--------------------------------------------------------------
where summation is carried out over index $i$ only, but not over $\alpha$.
The time derivative is already omitted, because we will discuss the
statistically stationary case. With substitution (\ref{substitution}), 
inequality (\ref{vol}) becomes
%--------------------------------------------------------------  
\begin{eqnarray}  
\label{vol1}  
\frac{H(\Gamma_r)}{V_r}
\le\sqrt{1+\frac{3}{4\pi r\kappa} \int_{B_r}\,
\left[\left(-u_i\partial_i+\kappa\partial_i^2\right)
\frac{\tilde{g}_{\alpha}^2}{2}
-\tilde{g}_{\alpha}(\partial_{\alpha}u_i) \tilde{g}_i 
-\tilde{g}_{\alpha}\partial_{\alpha}u_{\alpha} \tilde{G}\right]
\,\mbox{d}^3{\bf x}}\;. 
\end{eqnarray}  
%--------------------------------------------------------------  
We will consider now the four integrals under the square root separately
and denote them by  $I_1$, $I_2$, $I_3$, and 
$I_4$.
From (\ref{vol1}), with incompressibility, and by applying the Gauss theorem,  
it follows for $I_1$  
%--------------------------------------------------------------  
\begin{eqnarray}  
I_1=
%-\frac{3}{8\pi r \kappa}\int_{B_r}
%       {\bf \nabla}\cdot(\tilde{g}_{\alpha}^2 {\bf u})\,\mbox{d}^3{\bf x}
%\,,        
%            \nonumber \\
%\frac{3r}{2\kappa A_r}\oint_{\partial B_r}
%                      \tilde{g}_{\alpha}^2\,  
%                      {\bf u}\cdot\mbox{d}{\bf A}
\frac{3r}{2\kappa A_r}\oint_{\partial B_r}
                      \tilde{g}_{\alpha}^2
                      ({\bf u}-{\bf u}_0)\cdot\mbox{d}{\bf A}\,.
\end{eqnarray}
%--------------------------------------------------------------
$A_r=4\pi r^2$ is the surface content of $B_r$ and
the surface normal vector points toward the origin of $B_r$.
It is possible to add ${\bf u}_0={\bf u}({\bf x}_0)$,
the velocity at the center of $B_r$ for which 
$\langle {\bf u}_0\rangle_{\partial B_r}=0$. At this point, it has to 
be assumed that the fluctuations $\tilde{g}_{\alpha}^2$ 
are equally distributed over the sphere $\partial B_r$ in order to get 
$\oint \tilde{g}_{\alpha}^2\,{\bf u}_0\cdot\mbox{d}{\bf A}=0$.
The application of the Cauchy-Schwartz inequality results in
%--------------------------------------------------------------
\begin{eqnarray}
I_1&\le& \frac{3r}{2\kappa}
          \sqrt{\oint_{\partial B_r}
          \tilde{g}_{\alpha}^4\frac{\mbox{d}A}{A_r}}\,
          \sqrt{\oint_{\partial B_r}
          (({\bf u}-{\bf u}_0)\cdot{\bf n})^2\frac{\mbox{d}A}{A_r}}\,. 
\end{eqnarray}  
%--------------------------------------------------------------
The first square root is a scale resolved scalar gradient flatness which will be
discussed later and denoted by $F_4(r)=\langle g_{\alpha}^4\rangle_{B_r}/\langle
g_{\alpha}^2\rangle_V^2$.  Note that the $r$-dependence comes in via coarse
graining over balls of varying radius $r$.
The second term stands for the second order longitudinal velocity
structure function, which is $S_{\parallel}(r)=\epsilon r^2 /(15\nu)$ in the
viscous range, i.e.  on scales around and below
the Kolmogorov dissipation scale $\eta$.
We find
%--------------------------------------------------------------  
\begin{equation}  
\label{i1in} 
I_1\le\frac{3}{2\kappa}\sqrt{\frac{\epsilon F_4(r)}{15 \nu}}\,r^2\;.
\end{equation}  
%--------------------------------------------------------------

Integral $I_2$ in (\ref{vol1}) can be written as
%--------------------------------------------------------------  
\begin{eqnarray} 
\label{app1} 
I_2=\frac{3}{4\pi r}\int_{B_r}\,\left[\,\tilde{g}_{\alpha}{\bf\nabla}^2\tilde{g}_{\alpha}+
       |{\bf \nabla}\tilde{g}_{\alpha}|^2\,\right] \,\mbox{d}^3{\bf x}\,.               
\end{eqnarray}  
%-------------------------------------------------------------- 
Green's formula for scalars $u({\bf x})$ and $v({\bf x})$,
%--------------------------------------------------------------  
\begin{eqnarray}  
\int_V u({\bf x}) {\bf\nabla}^2 v({\bf x})\,\mbox{d}^3{\bf x}&=&
\oint_{\partial V} u({\bf x}) {\bf\nabla} v({\bf x})\cdot
\mbox{d}{\bf A}\nonumber\\
&-&\int_V {\bf\nabla} u({\bf x})\cdot {\bf\nabla} v({\bf x})\,
\mbox{d}^3{\bf x}\,,
\end{eqnarray}  
%-------------------------------------------------------------- 
with $u({\bf x})=v({\bf x})=\tilde{g}_{\alpha}({\bf x})$ is taken.
Substitution into (\ref{app1})
and application of the Cauchy-Schwartz inequality result in
%--------------------------------------------------------------  
\begin{eqnarray} 
I_2&=& \frac{3}{4\pi r}\oint_{\partial B_r}\,
       \tilde{g}_{\alpha}\,|{\bf \nabla}\tilde{g}_{\alpha}|\,{\bf n}
       \cdot\mbox{d}{\bf A}\nonumber\\ 
   &\le &3 r    
         \sqrt{\frac{1}{A_r}\oint_{\partial B_r}\tilde{g}_{\alpha}^2
         \,\mbox{d}A}   
          \;  
         \sqrt{\frac{1}{A_r}\oint_{\partial B_r}
         |{\bf \nabla}\tilde{g}_{\alpha}|^2\,\mbox{d}A}
         \nonumber\\ 
   &=& 3 r \sqrt{F_2(r)}\,\sqrt{\langle 
         |{\bf\nabla}\tilde{g}_{\alpha}|^2\rangle_{\partial B_r}}\,, 
\end{eqnarray}  
%-------------------------------------------------------------- 
with $F_2(r)=\langle g_{\alpha}^2\rangle_{\partial B_r}/ \langle
g_{\alpha}^2\rangle_V$ and 
${\bf \nabla}\tilde{g}_{\alpha}=|{\bf \nabla}\tilde{g}_{\alpha}|\,{\bf n}$.  We
assume that the volume average as well as the surface average
give the same results over scales $r$ due to homogeneity of turbulence.

The third integral, $I_3$, can be estimated by the Cauchy-Schwartz inequality to
%--------------------------------------------------------------  
\begin{eqnarray} 
I_3&=&-\frac{3}{4\pi r\kappa}\int_{B_r}\,\tilde{g}_{\alpha} (\partial_{\alpha} u_i)
       \tilde{g}_i\,\mbox{d}^3{\bf x}\nonumber\\
   &\le&\frac{r^2}{\kappa}\sum_{i=1}^3
   \sqrt{\langle \tilde{g}_{\alpha}^2 \tilde{g}_i^2 \rangle_{B_r}\, 
         \langle (\partial_{\alpha} u_i)^2\rangle_{B_r}}\,.    
\end{eqnarray}  
%-------------------------------------------------------------- 
It is reasonable to assume that the fluctuations along the mean scalar gradient
are the largest so that the mixed terms can be estimated by $\langle
\tilde{g}_{\alpha}^2 \tilde{g}_i^2 \rangle_{B_r}\le \langle \tilde{g}_{\alpha}^4
\rangle_{B_r}=F_4(r)$ for $i\ne \alpha$.  The velocity gradient can be treated
by the energy dissipation averaged over balls $B_r$. Here we estimated 
$\sum^3_{i=1}\langle (\partial_{\alpha} u_i)^2\rangle_{B_r}
\le\langle\epsilon\rangle_{B_r}/\nu$.
The notation $\epsilon_r=\langle\epsilon\rangle_{B_r}$ is used for the 
following and consequently 
%--------------------------------------------------------------  
\begin{eqnarray} 
I_3&\le&\frac{r^2}{\kappa}
   \sqrt{\frac{F_4(r)\epsilon_r}{\nu}}\,.    
\end{eqnarray}  
%-------------------------------------------------------------- 
Similarily one can proceed for the integral $I_4$ 
%--------------------------------------------------------------  
\begin{eqnarray} 
I_4=-\frac{3}{4\pi r\kappa}\int_{B_r}\,\tilde{g}_{\alpha} 
(\partial_{\alpha} u_{\alpha}) 
       \,\tilde{G}
       \,\mbox{d}^3{\bf x}
%   &\le&\frac{r^2 \tilde{G}}{\kappa}
%   \sqrt{\langle \tilde{g}_{\alpha}^2 \rangle_{B_r}\, 
%         \langle (\partial_{\alpha} u_{\alpha})^2\rangle_{B_r}}\,,\nonumber\\
   \le\frac{r^2 \tilde{G}}{\kappa}\sqrt{\frac{F_2(r)\epsilon_r}{\nu}}\,.    
\end{eqnarray}  
%-------------------------------------------------------------- 
In summary we get 
%--------------------------------------------------------------  
\begin{eqnarray} 
\frac{H(\Gamma_r)}{V_r}
\le\sqrt{1+\frac{3 r^2}{2\kappa}\sqrt{\frac{\epsilon F_4(r)}{15 \nu}}+
 3 r \sqrt{F_2(r) \langle 
         |{\bf\nabla}\tilde{g}_{\alpha}|^2\rangle_{\partial B_r}}+
 \frac{  r^2}{\kappa} \sqrt{\frac{F_4(r)\epsilon_r}{\nu}} +
 \frac{  r^2 \tilde{G}}{\kappa} \sqrt{\frac{F_2(r)\epsilon_r}{\nu}}}\,.
\label{first}
\end{eqnarray}    
%--------------------------------------------------------------
Unfortunately left with a couple of
unknown terms in this expression. Further progress can be made only
by the use of dimensional arguments.
One can expect that the energy dissipation field fluctuates most strongly around
scale $\eta$ and large velocity gradients will be smoothed
by the finite viscosity on scales below. Therefore it is reasonable to assume
%--------------------------------------------------------------
\begin{eqnarray}
\frac{\epsilon_r}{\epsilon}\le\frac{\epsilon_{\eta}}{\epsilon}
\simeq\left(\frac{\eta}{L}\right)^{\gamma-1}=
\left(\frac{20}{3}R_{\lambda}^{-2}\right)^{\frac{3}{4}(\gamma-1)}\,,
\end{eqnarray}
%--------------------------------------------------------------
where $(\eta/L)=(20/3)^{3/4} R_{\lambda}^{-3/2}$ 
\cite{Pope00}.
$\gamma$ is a (universal) scaling exponent varying between $\gamma_{1}$ and
$\gamma_{2}$.  Clearly, the minimum exponent, $\gamma_{1}$, will be the 
dominant one.

The following term in $I_2$,
can be simplified to
%--------------------------------------------------------------  
\begin{eqnarray} 
\langle|{\bf\nabla}\tilde{g}_{\alpha}|^2\rangle_{\partial B_r}\le
\frac{1}{\eta_B^2}\langle(\tilde{g}_{\alpha})^2\rangle_{\partial B_r}=
\frac{F_2(r)}{\eta_B^2}\,,
\end{eqnarray}
%--------------------------------------------------------------   
and $F_2(r)$ follows in lines with (17) to
%--------------------------------------------------------------
\begin{eqnarray}
F_2(r)=\frac{\langle g_{\alpha}^2\rangle_{\partial B_r}}
            {\langle g_{\alpha}^2\rangle_V}
            \simeq
       \frac{\epsilon_{\theta, r}}
            {\epsilon_{\theta}}
            \le
       \frac{\epsilon_{\theta, \eta_B}}
            {\epsilon_{\theta}}
            \simeq
       \left(\frac{\eta_B}{L}\right)^{\delta-1}\;,
\end{eqnarray}
%--------------------------------------------------------------
where $\epsilon_{\theta, r}= \langle\epsilon_{\theta}\rangle_{\partial B_r}$
is the coarse grained scalar dissipation rate.
With $\eta_B/\eta=Sc^{-1/2}$ one gets
%--------------------------------------------------------------
\begin{eqnarray}
F_2(r)\le \left[
          \left(
          \frac{20}{3}
          \right)^{\frac{3}{4}}
          Sc^{-\frac{1}{2}} R_{\lambda}^{-\frac{3}{2}}
          \right]^{\delta-1} \;.
\end{eqnarray}
%--------------------------------------------------------------
Again the scaling exponent $\delta \in [\delta_1,\delta_2]$.
$F_4(r)$ can be estimated as follows
%--------------------------------------------------------------
\begin{eqnarray}
F_4(r)=\frac{\langle g_{\alpha}^4\rangle_{B_r}}
            {\langle g_{\alpha}^2\rangle_V^2}
      =\frac{\langle g_{\alpha}^4\rangle_{B_r}}
            {\langle g_{\alpha}^4\rangle_V} F_4(L)
            \le
       \left(\frac{L}{\eta_B}\right)^{4} F_4(L)
\end{eqnarray}
%--------------------------------------------------------------
where 
$F_4(L)=\langle g_{\alpha}^4\rangle_V/\langle g_{\alpha}^2\rangle_V^2$,
$\langle g_{\alpha}^4\rangle_{B_r}\le G^4/\eta_B^4$, and 
$\langle g_{\alpha}^4\rangle_{V}\simeq G^4/L^4$.
$G$ is the magnitude of the mean scalar gradient. 
It is also assumed that scalar gradient moments are at maximum
around the Batchelor scale.
Thus follows
%--------------------------------------------------------------
\begin{eqnarray}
F_4(r)\le \frac{27}{8000} Sc^2 R_{\lambda}^6 F_4(L)\,.
\end{eqnarray}
%--------------------------------------------------------------
The term $I_3$ will be much larger than $I_4$ if for the mean scalar gradient 
holds, $\tilde{G}\ll \sqrt{F_4(r)/F_2(r)}$, as it will be the case for the
comparison with numerical experiments. 
Finally, $I_1$ will always be subdominant compared to $I_3$
due to $\epsilon_r\gg \epsilon$.

All the estimates are inserted now into (\ref{first}) and
we assume a scaling relation for the relative Hausdorff volume,
$H(\Gamma_r)/V_r\sim r^{D_H-3}$. Additionally, all scales are expressed in units
of Kolmogorov dissipation length, $\tilde{r}=r/\eta$.
The Hausdorff dimension of the graph over two-dimensional balls is
given by the additive law, $D^{\prime}_H=D_H-1$, \cite{Mattila75} when 
assuming almost isotropic graphs,
%--------------------------------------------------------------
\begin{eqnarray}
D^{\prime}_H
&=&2+\frac{\mbox{d}\,\ln  (H(\Gamma_{\tilde{r}})/V_{\tilde{r}})}
        {\mbox{d}\,\ln \tilde{r}}\nonumber\\
&\le& 2+\frac{\mbox{d}}{\mbox{d}\,\ln \tilde{r}}\,\ln
\sqrt{1+
      3\left[
       \left(\frac{20}{3}\right)^{\frac{3}{4}}
       R_{\lambda}^{-\frac{3}{2}}
       \right]^{\delta_1-1} Sc^{\frac{2-\delta_1}{2}}\,\tilde{r} +
\sqrt{\frac{27 F_4(L)}{8000}}\, 
\left(\frac{20}{3}\right)^{\frac{3}{8}(\gamma_1-1)}
Sc^2\, R_{\lambda}^{\frac{15-3\gamma_1}{4}}\,\tilde{r}^2}\,.
\label{final}
\end{eqnarray}
%--------------------------------------------------------------
The expression probes the local slope of the scaling relation. The upper bound
of $D^{\prime}_H$ will be determined by the leading term under the square
root at every scale $\tilde{r}$ and we expect the result 
$2\le D_H^{\prime}\le 3$. 

The missing scalar derivative flatness factor $F_4(L)$ can be evaluated from
\cite{PK02} and present simulations.  The latter are conducted in a
homogeneously sheared flow in which the scalar field of constant gradient was
allowed to evolve according to the advection-diffusion equation \cite{Schu03,Schu00}.
The mean scalar gradient is kept the same in all runs 
resulting in dimensionless $\tilde{G}=0.37$ at $Sc=1$ and 0.05 at $Sc=64$
so that $\tilde{G}\ll \sqrt{F_4(r)/F_2(r)}$ is justified.  
As the inset of
Fig.~2 shows, the flatness factors for both simulations are $\sim 10$ for
$Sc>1$ and will be taken constant for the following.

The two minimum scaling exponents, $\gamma_1$ and $\delta_1$, can be taken from
experiments where the whole multifractal spectrum for $\epsilon_{\theta}({\bf
x},t)$ and $\epsilon({\bf x},t)$ was measured, respectively.  Meneveau and
Sreenivasan \cite{Meneveau} found for experimental data at different Reynolds
numbers that $\gamma_1$ is about 1/4.  Prasad {\it et al.}  found a value of
$\delta_{1}\approx 2/5$ in a high-Schmidt number scalar mixing experiment at
$Sc\approx 2000$ \cite{Prasad}.

Figure 2 plots (\ref{final}) for three different values of $Sc$.  For scales
accessible to the simulations, i.e, larger than $\eta_B$ the upper bound is
found to be always 3.  While for the scalar case the crossover of $D^{\prime}_H$
from 2 to 3 takes place around $\eta_B$ \cite{GroLoh94}, it is present here for
by far smaller scales.  It is just the fairly rough estimate for the additional
scalar gradient stretching term (last term on the l.h.s.  of (4)) that causes
the crossover scale, $\tilde{r}_c$, to be smaller than $\eta_B$.  When inserting
the numbers, one gets $\tilde{r}_c\sim Sc^{-6/5} R_{\lambda}^{-213/80}$, at
which the second term of (23) starts to dominate.  To illustrate this effect, we
kept $I_1$ only and plot the corresponding $D_H^{\prime}$ as dashed lines in
Fig.~2.  The bounds are shifted then by an order of magnitude, but still rough
due to estimate (22).

For comparison, we conducted the relative Hausdorff volume of the scalar gradient
field from numerical simulation data.  Two-dimensional ``balls'' are then
squares of a dyadic grid of sidelength $2^{-j}\times 2\pi$.  The corresponding
local slope of the scaling dimension is added in Fig.~2 and a plateau with a
non-integer dimension of $D_H^{\prime}\approx 2.3$ can be observed for
intermediate scales.  This feature cannot be reproduced by the upper bounds
because the $\tilde{r}$-dependence of all four $I$ terms has integer powers
only. Even for passive scalars, non-integer $D_H$ were obtained only for
inertial range scaling of the velocity structure function of $\sim
r^{2/3}$ (cf.~\cite{GroLoh94}).

To summarize, we have discussed the applicability of the geometric measure 
theory to scalar gradient fields for high-$Sc$ mixing. The upper bounds
on $D_H$ are consistent with simulation results but rough.
Additionally, they are insensitive to the particular choice
of the threshold value of the level set and thus cannot capture slight
differences between positive and negative level sets of same magnitude which
were discussed in \cite{Schu03}.  Interestingly, a dependence of the bounds on
the Taylor-Reynolds number results.  Batchelor's original model for the
visocous-convective range was thought to be insensitive to the physics in the
inertial range, i.e., on scales larger $\eta$ \cite{Batchelor}.  A detailed
investigation of that problem will be a part of future work and further 
improvement and extension of the bounds might be possible therefore.

The numerical computations were carried out on the IBM Blue Horizon at the San
Diego Supercomputer Center within the NPACI program of the US National Science
Foundation which we wish to acknowledge.  Comments by C.~R.~Doering,
B.~Eckhardt, and K.~R.~Sreenivasan are acknowledged.

%------References------------------------------------------  

%-------------------------------------------------------------------------------
\begin{figure}
\begin{center}
\epsfig{file=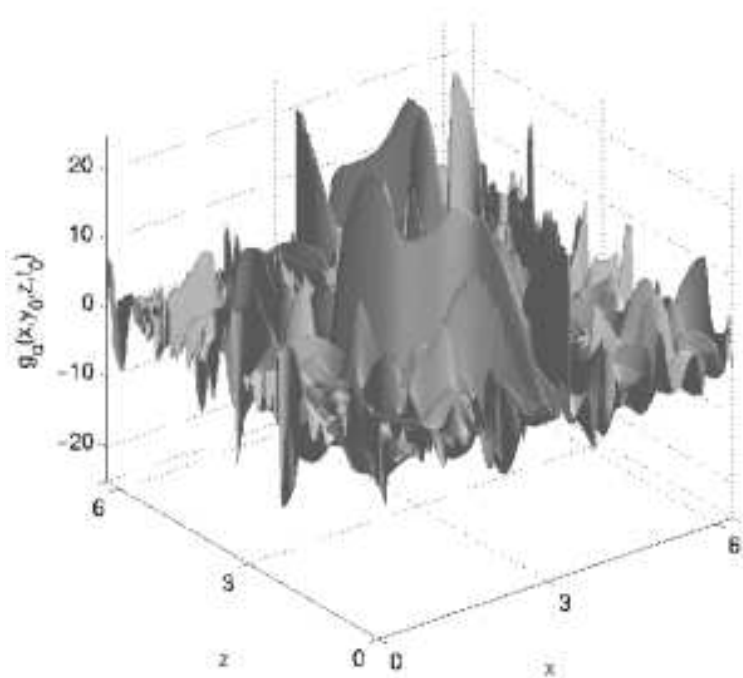,width=10cm}
\end{center}
\caption{The graph of the scalar gradient field $g_{\alpha}$ which is shown here
over the $x-z$ plane at fixed $y_0$ and $t_0$ for numerical data at $Sc=16$. Grid
resolution was $N=512$.}
\label{fig1}
\end{figure}
%-------------------------------------------------------------------------------
\begin{figure}
\begin{center}
\epsfig{file=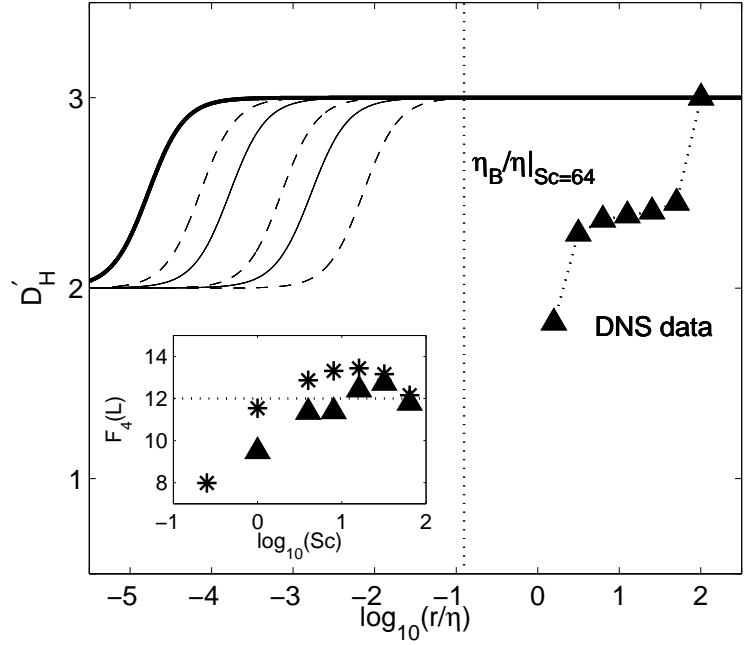,width=10cm}
\end{center}
\caption{Scale dependent upper bound of the Hausdorff dimension of scalar
gradient level sets for different Schmidt numbers. Value of $R_{\lambda}=87$
is taken from [4].  Solid
lines are for $Sc=64$ (thick line), 6.4, and 0.64 from left to right.  Dashed
lines are for bounds at the same $Sc$ if integral $I_1$ would dominate in (23)
as in the scalar case [8].  The Batchelor scale for $Sc=64$ is
marked as a dotted vertical line.  The panel contains also the DNS findings
for the scale resolved $D_H^{\prime}$ for $Sc=64$ with a plateau at a
value of about 2.3 .  Inset:  Flatness factor $F_4(L)$ of the scalar derivative
along the mean scalar gradient, $g_{\alpha}$, as a function of the Schmidt
number.  The dotted line stands for the mean of all data points and is drawn at
12.  Present data are indicated by triangles and data from [1] by
asterisks.}
\label{fig2}
\end{figure}
\end{document}